\documentclass[a4paper,11pt]{article}

\usepackage{lineno}
\usepackage{ulem}

\usepackage{verbatim} 
\usepackage{booktabs}

\usepackage{ifthen}
\newboolean{pdflatex}
\setboolean{pdflatex}{true} 

\newboolean{articletitles}
\setboolean{articletitles}{true} 

\newboolean{uprightparticles}
\setboolean{uprightparticles}{false} 

\newboolean{inbibliography}
\setboolean{inbibliography}{false} 

\usepackage{xspace} 
\usepackage{upgreek}

\def\MagUp {\mbox{\em Mag\kern -0.05em Up}\xspace}

\ifthenelse{\boolean{uprightparticles}}%
{

 \def\PDelta      {\ensuremath{\Delta}\xspace}                 
 \def\PXi      {\ensuremath{\Xi}\xspace}                 
 \def\PLambda      {\ensuremath{\Lambda}\xspace}                 
 \def\PSigma      {\ensuremath{\Sigma}\xspace}                 
 \def\POmega      {\ensuremath{\Omega}\xspace}                 
 \def\PUpsilon      {\ensuremath{\Upsilon}\xspace}                 
 
 \def\PB      {\ensuremath{\mathrm{B}}\xspace}                 
                  
 \def\PD      {\ensuremath{\mathrm{D}}\xspace}

 \def\PK      {\ensuremath{\mathrm{K}}\xspace}

 \def\PZ      {\ensuremath{\mathrm{Z}}\xspace}

 \def\Pi      {\ensuremath{\mathrm{i}}\xspace}

 \def\Pq      {\ensuremath{\mathrm{q}}\xspace}

}
{

 \mathchardef\PDelta="7101
 \mathchardef\PXi="7104
 \mathchardef\PLambda="7103
 \mathchardef\PSigma="7106
 \mathchardef\POmega="710A
 \mathchardef\PUpsilon="7107
                  
 \def\PB      {\ensuremath{B}\xspace}                 
                  
 \def\PD      {\ensuremath{D}\xspace}

 \def\PK      {\ensuremath{K}\xspace}

 \def\PZ      {\ensuremath{Z}\xspace}

 \def\Pi      {\ensuremath{i}\xspace}

 \def\Pq      {\ensuremath{q}\xspace}

}

\makeatletter
\ifcase \@ptsize \relax
  \newcommand{\miniscule}{\@setfontsize\miniscule{4}{5}}
\or
  \newcommand{\miniscule}{\@setfontsize\miniscule{5}{6}}
\or
  \newcommand{\miniscule}{\@setfontsize\miniscule{5}{6}}
\fi
\makeatother

\DeclareRobustCommand{\optbar}[1]{\shortstack{{\miniscule (\rule[.5ex]{1.25em}{.18mm})}
  \\ [-.7ex] $#1$}}

\def\Z      {{\ensuremath{\PZ}}\xspace}

\def\quark     {{\ensuremath{\Pq}}\xspace}
\def\quarkbar  {{\ensuremath{\overline \quark}}\xspace}

  \def\Kbar    {{\kern 0.2em\overline{\kern -0.2em \PK}{}}\xspace}

\def\KorKbar    {\kern 0.18em\optbar{\kern -0.18em K}{}\xspace}

  \def\Dbar    {{\kern 0.2em\overline{\kern -0.2em \PD}{}}\xspace}

\def\DorDbar    {\kern 0.18em\optbar{\kern -0.18em D}{}\xspace}

\def\Bbar    {{\ensuremath{\kern 0.18em\overline{\kern -0.18em \PB}{}}}\xspace}

\def\BorBbar    {\kern 0.18em\optbar{\kern -0.18em B}{}\xspace}

  \def\Y#1S{\ensuremath{\PUpsilon{(#1S)}}\xspace}

\def\Lbar        {{\ensuremath{\kern 0.1em\overline{\kern -0.1em\PLambda}}}\xspace}
\def\LorLbar    {\kern 0.18em\optbar{\kern -0.18em \PLambda}{}\xspace}

\newcommand{\decay}[2]{\ensuremath{#1\!\to #2}\xspace}         

\def\to                 {\ensuremath{\rightarrow}\xspace}

\def\AT#1     {\ensuremath{A_{\mathrm{T}}^{#1}}\xspace}           

\def\C#1      {\ensuremath{\mathcal{C}_{#1}}\xspace}                       
\def\Cp#1     {\ensuremath{\mathcal{C}_{#1}^{'}}\xspace}                    
\def\Ceff#1   {\ensuremath{\mathcal{C}_{#1}^{\mathrm{(eff)}}}\xspace}        
\def\Cpeff#1  {\ensuremath{\mathcal{C}_{#1}^{'\mathrm{(eff)}}}\xspace}       
\def\Ope#1    {\ensuremath{\mathcal{O}_{#1}}\xspace}                       
\def\Opep#1   {\ensuremath{\mathcal{O}_{#1}^{'}}\xspace}                    

\newcommand{\unit}[1]{\ensuremath{\mathrm{ \,#1}}\xspace}          

\newcommand{\tev}{\ifthenelse{\boolean{inbibliography}}{\ensuremath{~T\kern -0.05em eV}\xspace}{\ensuremath{\mathrm{\,Te\kern -0.1em V}}}\xspace}
\newcommand{\gev}{\ensuremath{\mathrm{\,Ge\kern -0.1em V}}\xspace}
\newcommand{\mev}{\ensuremath{\mathrm{\,Me\kern -0.1em V}}\xspace}
\newcommand{\kev}{\ensuremath{\mathrm{\,ke\kern -0.1em V}}\xspace}
\newcommand{\ev}{\ensuremath{\mathrm{\,e\kern -0.1em V}}\xspace}
\newcommand{\gevc}{\ensuremath{{\mathrm{\,Ge\kern -0.1em V\!/}c}}\xspace}
\newcommand{\mevc}{\ensuremath{{\mathrm{\,Me\kern -0.1em V\!/}c}}\xspace}
\newcommand{\gevcc}{\ensuremath{{\mathrm{\,Ge\kern -0.1em V\!/}c^2}}\xspace}
\newcommand{\gevgevcccc}{\ensuremath{{\mathrm{\,Ge\kern -0.1em V^2\!/}c^4}}\xspace}
\newcommand{\mevcc}{\ensuremath{{\mathrm{\,Me\kern -0.1em V\!/}c^2}}\xspace}

\def\m    {\ensuremath{\mathrm{ \,m}}\xspace}

\def\cm   {\ensuremath{\mathrm{ \,cm}}\xspace}

\def\mm   {\ensuremath{\mathrm{ \,mm}}\xspace}

\def\mum  {\ensuremath{{\,\upmu\mathrm{m}}}\xspace}

\def\nb {\ensuremath{\mathrm{ \,nb}}\xspace}

\def\fb   {\ensuremath{\mbox{\,fb}}\xspace}

\def\gsim{{~\raise.15em\hbox{$>$}\kern-.85em
          \lower.35em\hbox{$\sim$}~}\xspace}
\def\lsim{{~\raise.15em\hbox{$<$}\kern-.85em
          \lower.35em\hbox{$\sim$}~}\xspace}

\def\geant      {\mbox{\textsc{Geant4}}\xspace}
\def\mokka      {\mbox{\textsc{Mokka}}\xspace}

\def\tell1  {TELL1\xspace}
\def\ukl1   {UKL1\xspace}

\newcommand{\cepc}{CEPC\xspace}
\newcommand{\cepcZpole}{\cepc\ \Z pole\xspace}

\newcommand{\fccee}{\mbox{FCC-ee}\xspace}
\newcommand{\hit}{\mathrm{hit}}
\newcommand{\charge}{\mathrm{charge}}
\newcommand{\ZToqq}{\decay{\Z}{\quark\quarkbar}}
\newcommand{\cmsec}{\ensuremath{\mathrm{{\cm^2\,s^{-1}}}}\xspace}
\newcommand{\E}[1]{\ensuremath{\times 10^{#1}}\xspace}
\newcommand{\kb}{\ensuremath{k}\xspace}
\newcommand{\vion}{\ensuremath{V_{\mathrm{ion}}}\xspace}

\pdfoutput=1 

\usepackage{jinstpub} 

\title{\boldmath Feasibility study of TPC at electron positron colliders at Z
pole operation}

\author[a,c]{M. Zhao}
\author[b,1]{, M. Ruan\note{Corresponding author.}}
\author[b]{, H. Qi}
\author[a]{, Y. Gao}

\affiliation[a]{Center of High Energy Physics, Tsinghua University,\\ Beijing 100084, China}
\affiliation[b]{Institute of High Energy Physics, Chinese Academy of Sciences,\\ Beijing 100049, China}
\affiliation[c]{Department of Nuclear Physics, China Institute of Atomic Energy,\\ Beijing 102413, China}
\emailAdd{manqi.ruan@ihep.ac.cn}

\abstract{TPC is a promising technology for the future electron positron colliders. However, its application might be limited at high event rate and high hit occupancies. In this paper, we study the feasibility of using TPC at the circular electron positron collider(\cepc) at \Z pole using full simulated \ZToqq samples. By evaluating the local charge density and voxel occupancy at different TPC parameters. Our study shows that the TPC could be applied to the \cepcZpole operation if back flow ion is controlled to per mille level. We also suggest the applicable TPC parameters for \fccee\ \Z pole operations, whose instant luminosity is $2\E{36} \cmsec$, 2 orders of magnitude higher than that of \cepc.}

\collaboration[c]{on behalf of CEPC collaboration}

\begin{document}
\maketitle
\flushbottom

\section{Introduction}
\label{sec:introduction}
The CEPC is a proposed electron positron collider after the Higgs discovery. It will be applied as a Higgs factory and \Z factory. 
As a Higgs factory, it will be operated at $240 \unit{GeV}$ center of mass energy, produce 1 million Higgs bosons in 10 years and measure the Higgs couplings to $0.1\% - 1\%$ level accuracy\cite{ahmad2015cepc}. It will also be operated at the \Z pole and produce approximately 10 billion \Z bosons each year. The typical cross-sections and event rates for nominal \cepc\ accelerator parameters are given in Table~\ref{tab:accelerator parameters}. 
\begin{table}[h]
    \label{tab:accelerator parameters}
    \caption{\small
        Typical parameters and productivity of the CEPC
    }
    \begin{center}
        \begin{tabular}{ccc}
            \toprule
            & Higgs runs & \Z pole \\
            \hline
            Center of mass energy(GeV)  & 240 & 91 \\
            Instant luminosity($\cmsec$) & $2\E{34}$&$2\E{34}$ \\ 
            Signal cross-section &$200\fb$ & $30\nb$ for \ZToqq\\ 
            \bottomrule
        \end{tabular}
    \end{center}
\end{table}

TPC has been widely used in high energy physics experiments\cite{behnke2007ilc,atwood1991performance,anderson2003star,nappi2000alice}. It provide high-efficiency track finding, precise momentum measurement, and has low material budget. In addition, TPC provides good $\mathrm{d}E/\mathrm{d}x$ measurement, providing essential information for electron identification and particle identification. These benefits are highly appreciated for the \cepc\ physics program\cite{ahmad2015cepc}.

The \cepc\ conceptual detector is a PFA oriented detector designed following the ILD detector\cite{behnke2007ilc}, where TPC is used as the main tracker. The geometry of the CEPC conceptual detector is adjusted to the \cepc\ collision circumstance. At a center of mass energy close to the \Z mass, most of the TPC hits are induced by \ZToqq events. The event rate of \ZToqq process is $600 \unit{Hz}$ per IP for the \cepcZpole operation\cite{ahmad2015cepc}. Another future electron positron collider, the \fccee\cite{wenninger2014future}, has much aggressive beam parameters and the event rate of \ZToqq is $60\unit{kHz}$, two orders of magnitude higher than that of CEPC. Such a high event rate makes stringent requirement for the TPC. It is crucial to study the feasibility of using TPC at these electron positron colliders at their \Z pole operation. 

On the other hand, the response of TPC is slow comparing to the silicon detectors, as the electrons generated in the primary ionization need to shift toward the endcap to create electronic signals. The position measurement of TPC is limited by the back flow ions, i.e., ions generated in the amplification back flowing into the gas volume. These spatial ion charges will distort the electronic field and induces uncertainties to the hit position measurements. 

The performance of TPC is mainly limited by local charge density and the voxel occupancy. In this paper, we evaluate these effects with full-simulated \ZToqq data, extract the analytic format of corresponding distributions and parameterized them to TPC and luminosity parameters. Taking reference to the empirical formula on these parameters, we conclude that the TPC works at nominal instant luminosity at CEPC, if the charge induced by back flow ion is only 1 order of magnitude higher than that of primary ionization.

\section{Simulation and sample}
We simulated 9 thousand \ZToqq events with \mokka\cite{de2002detector}, the \geant simulation package\cite{agostinelli2003geant4} for CEPC detector optimization study. Giving a $2\times10^{34} \cmsec$ instant luminosity, this sample is corresponding to a data-taking period of 15 seconds.  

The detector geometry we used in simulation is the \cepc\ conceptual detector model, the conceptual detector modified from the ILD detector geometry. Designed following the principle of particle flow algorithm, the conceptual detector uses low material tracking system and high granularity calorimeter system. Both ECAL and HCAL are installed inside the solenoid magneto to reduce the dead zone. 

\cepc\ conceptual detector uses TPC as main tracker. The TPC has an inner radius of $330\mm$, an outer radius of $1808\mm$ and a half-length of $2350\mm$. The readout pixel is of $1\mm$ along the $\phi-$direction and $6\mm$ along the radial direction ($r-$direction). Along the $r-$direction, the TPC is divided into 220 layers. The working gas of the TPC is T2K gas at the pressure of 1 atm, which is composed of 95\% argon, 2\% isobutene and 3\% CF4.

\section{Spatial charge distribution in the TPC (Physics picture)}
\begin{figure}[tb]
    \begin{center}
        \includegraphics[width=0.60 \textwidth]{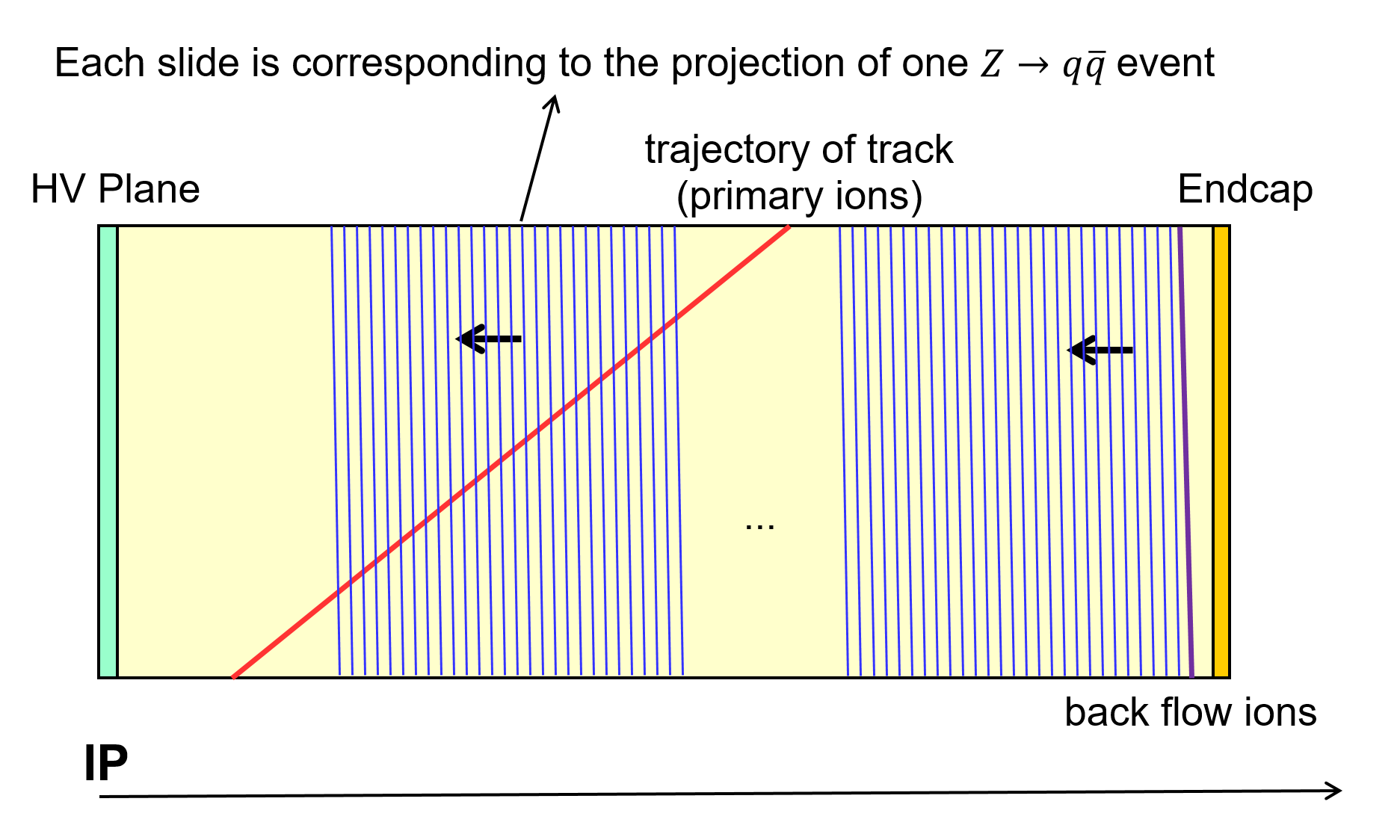}
        \vspace*{-0.5cm}
    \end{center}
    \caption{ \small
        A schematic one-forth view of TPC. The red line represents the trajectory of a track and the primary ions. The purple line in the right represents the back flow ions corresponding to the primary ions(red line). Each blue line represents the back flow ions corresponding to a \ZToqq event.  
    }
    \label{fig:ion}
\end{figure}

The ions in the TPC volume are induced by primary ionization and the ion back flow. Once a charged track is sailing through the TPC volume, it induces primary ionizations along its trajectory. Driven by the drifting electric field, the electrons are drifted to the endcap and the ions to the central HV plane. Once the electrons arrive the endcap, they creates electronic signals through the cascade amplification with a typical gain of $3000 - 6000$, depending on different detector technology. A fraction of the ions generated in the amplification procedure could back flow into the TPC volume, inducing the back flow ions. Since the ion is much heavier than the electron, the shift velocity of ion is typically four orders of magnitude smaller than that of the electron. Therefore, for a given track, the ions generated in primary ionizations are located along the track helix, while the back flow ions forms effectively a projection of the initial track to the transverse plane, as shown in Figure~\ref{fig:ion}.

At the ILD TPC, the primary charge density would be of around $100\unit{P.I\,cm^{-1}}$ along the track\cite{sauli1977principles}).
The shift velocity is $80\unit{\,km\,s^{-1}}$ for the electron and $5 \unit{\,m\,s^{-1}}$ for the ions. Giving the benchmark luminosity of $2 \times 10^{34} \cmsec$ at \cepc\ \Z pole operation, the \ZToqq events have an event rate of $600\unit{Hz}$. Since the TPC half $z$ is $2.35\m$, there will be roughly $600$ disks located evenly in the TPC volume, i.e, $300$ disks at each side of the HV plane, each corresponding to the back flow ions induced by one \ZToqq event, as shown in Figure~\ref{fig:ion}. These disks are then shifted toward the central HV plane with the velocity of ions. 

These ions distributed could be described by projective charge density and local hit density. The latter is described by the voxel occupancy. In the following part of this paper, we will resolve the spatial distribution of the hits and charges, from which we calculated the distortion at different parameters configuration. 

\section{Hit map and voxel occupancy}

\begin{figure}[tb]
    \begin{center}
        \includegraphics[width=0.60 \textwidth]{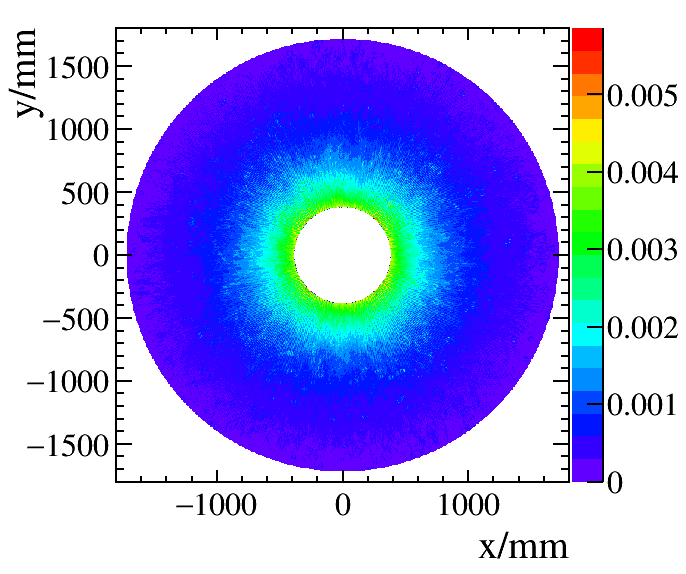}
        \vspace*{-0.5cm}
    \end{center}
    \caption{ \small
        Projective hit map on the X-Y plane for \ZToqq events. The $z$ axis represents the hit density in unit $\mm^{-2}$ and it is normalized to one \ZToqq event.  
    }
    \label{fig:hits map}
\end{figure}

The projective TPC hit map of $9$ thousand \ZToqq events is shown in Figure~\ref{fig:hits map}, which exhibits an uniform distribution along the $\phi$-direction. The hit density decreases with increasing the radius. 

\begin{figure}[tb]
    \begin{center}
        \includegraphics[width=0.60 \textwidth]{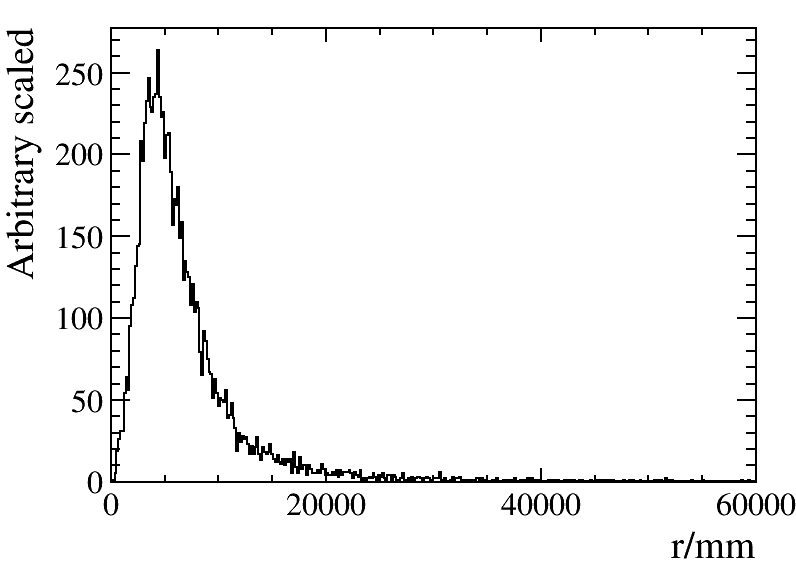}
        \vspace*{-0.5cm}
    \end{center}
    \caption{ \small
        Number of hits per \ZToqq event distribution.
    }
    \label{fig:number of hits}
\end{figure}

The distribution of hits per event is given in Figure~\ref{fig:number of hits}. 60 million hits are generated in this sample. On average, each \ZToqq event will induce $6900$ hits in the TPC volume and the most probably value of the number of hits is about $4000$.

The local hit density could be resolved from the hit map as shown in Figure~\ref{fig:hits map}. For 9 thousand \ZToqq events, the average hit density is $6 \unit{hits\mm^{-2}}$ on the transverse plane, and the peak value is 6 times larger than the average value, corresponding to the inner most layer of TPC. The dependence of local hit density and the radius is then extracted from fit. The fit function used is 
\begin{equation}
    \rho_{\hit} = \frac{A_{\hit}}{r-r_{0,\hit}} + \rho_{0,\hit}
\end{equation}. 
The data sample is normalized to the hit density in one \ZToqq event in half TPC.
The fit result can be found in Figure~\ref{fig:density}.

The voxel occupancy is the number of voxels which see a signal, divided
by all voxels in the TPC. voxel takes the configuration of TPC pixel, is 1 mm along the $\phi-$direction and 6 mm along the $r-$direction. The voxel size along $z-$direction is the electron shift speed divided by the DAQ rate, which gives 2 mm.
Therefore, an voxel has a size of $1\unit{mm} \times 6\unit{mm} \times 2\unit{mm} = 12\unit{mm^{3}}$. For each second, the number of voxels is the number of readout pixel in the end cap multiple by the DAQ rate. Assume the DAQ works at $40 \unit{MHz}$, the number of voxels for 1 second is $1.33 \times 10^{14}$.

At CEPC benchmark luminosity, 600 \ZToqq events will be generated in 1 second and 4 million TPC hits will be induced by these events. The beam bunches of CEPC are distributed evenly along the tunnel. The average voxel occupancy is then $3\times10^{-8}$. 

The voxel occupancy is also a function of the radius and it should simply follows the distribution of local hit density, as shown in Figure~\ref{fig:density}. Given the fact that the peaking hit density is 6 times larger than the average density, the maximal voxel occupancy is located at the TPC inner most layer, corresponding to a value of $2\times10^{-7}$. 

The voxel occupancy is proportional to the instant luminosity. Therefore, at the \fccee\ benchmark luminosity, the maximal voxel occupancy will be increased by 2 orders of magnitude, reaching the level of $2\times10^{-5}$. 

The CEPC proposes also the partial double ring design, where bunches are zipped into bunch trains with the typical length of 1km, two orders of magnitude smaller than the accelerator circumference. In this case, the voxel occupancy would also be increased by two orders of magnitude, reaching $2\times10^{-5}$ at the inner most layer. 

To conclude, for the CEPC \Z pole runs with the TPC of the conceptual detector, the voxel occupancy takes its maximal value between $2\E{-5}$ to $2\E{-7}$, which is safety for the \Z pole operation. 

\section{Projective charge density and constrains to the TPC parameters}
\begin{figure}[tb]
    \begin{center}
        \includegraphics[width=0.49 \textwidth]{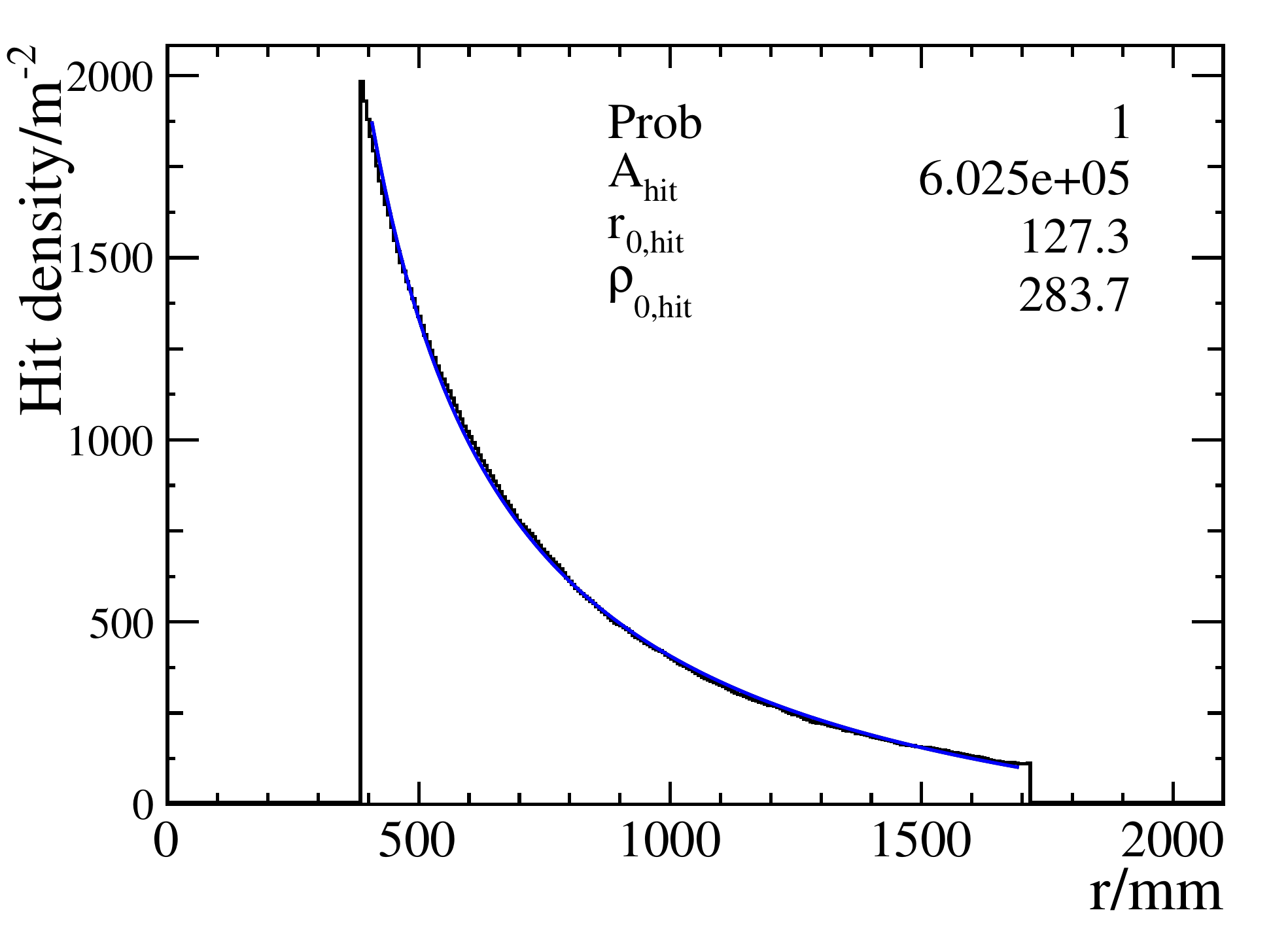}
        \includegraphics[width=0.49 \textwidth]{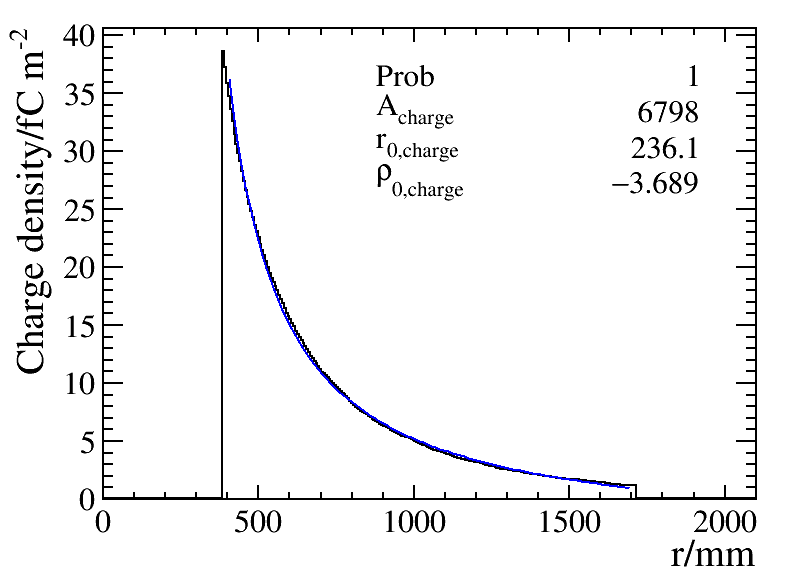}
        \vspace*{-0.5cm}
    \end{center}
    \caption{ \small
        Hit density(left) and charge density(right) as a function of radius. The distributions are normalized to one \ZToqq event.
    }
    \label{fig:density}
\end{figure}

\begin{figure}[tb]
    \begin{center}
        \includegraphics[width=0.60 \textwidth]{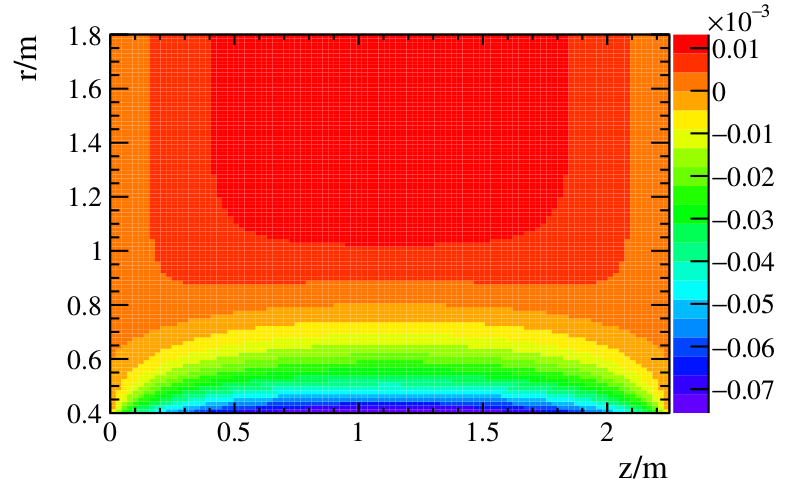}
        \vspace*{-0.5cm}
    \end{center}
    \caption{ \small
        One-forth view of $E_r/E_z$ in TPC with $L=2$ in the unit of $10^{34} \cmsec$, $\kb=5$ and $\vion=5$ in the unit of $\m\unit{\,s^{-1}}$.
    }
    \label{fig:EMField}
\end{figure}

The projective charge density could calculated from the local hit density weighted by the expected trajectory length and helix direction projection to the X-Y plane. The function
\begin{equation}
    \rho_{\charge} = \frac{A_{\charge}}{r-r_{0,\charge}} + \rho_{0,\charge}
\end{equation} is used to fit the data. The data sample is normalized to the charge density of primary ions in half TPC volume projection to the transverse plane. The fit results are shown in the right plot of Figure~\ref{fig:density}. 
The space charge density is proportional to the instant luminosity, the average number of back flow ions generated per primary ionization, and the inverse of the ion shift velocity. The latter determines how many ion disks can be hosted in the TPC volume. Extracting from the fit result, the charge density of all the ions could be expressed as:
\begin{equation}
    (1+k)\frac{L}{V_{\mathrm{ion}}}(\frac{4}{r-235}-0.0022)\unit{\,fC\,cm^{-3}},
\end{equation}
where $L$ is the luminosity normalized to $1\E{34} \cmsec$, $V_{\mathrm{ion}}$ is the ion drift velocity with unit in $\m\unit{\,s^{-1}}$. $r$ is the radius with unit in mm. These expressions are only an approximation valid with radius from $400\mm$ to $1700\mm$.  At the current TPC R\&D, the typical values of these parameters are:  
\begin{itemize}
    \item $\vion$: $5-10$.
    \item $\kb$: $10-100$ depending on the control of back flow ion.
\end{itemize}
These ion charges induce an electric field on top of the existing electromagnetic field. The electrons generated by the track ionization is drifted accordingly. Comparing to the case of zero ion charge density, the ion-induced electric field will cause an hit position distortion along $\phi-$direction.
The distortion in a local volume can be expressed as:
\begin{displaymath}
    \Delta l = \frac{\omega\tau}{1+(\omega\tau)^2} \times \frac{E_r}{E_z} \Delta z,
\end{displaymath}
where $\omega \equiv eB/m$ and $\tau$ is the mean free time of electrons. The value of $\omega \tau$ is quite large using T2K gas and it vary with gas pressure and electric magnetic field. The value $\omega\tau=10$ is used for estimation. The value of $E_r/E_z$ is shown in Figure~\ref{fig:EMField}.

\begin{figure}[tb]
    \begin{center}
        \includegraphics[width=0.60 \textwidth]{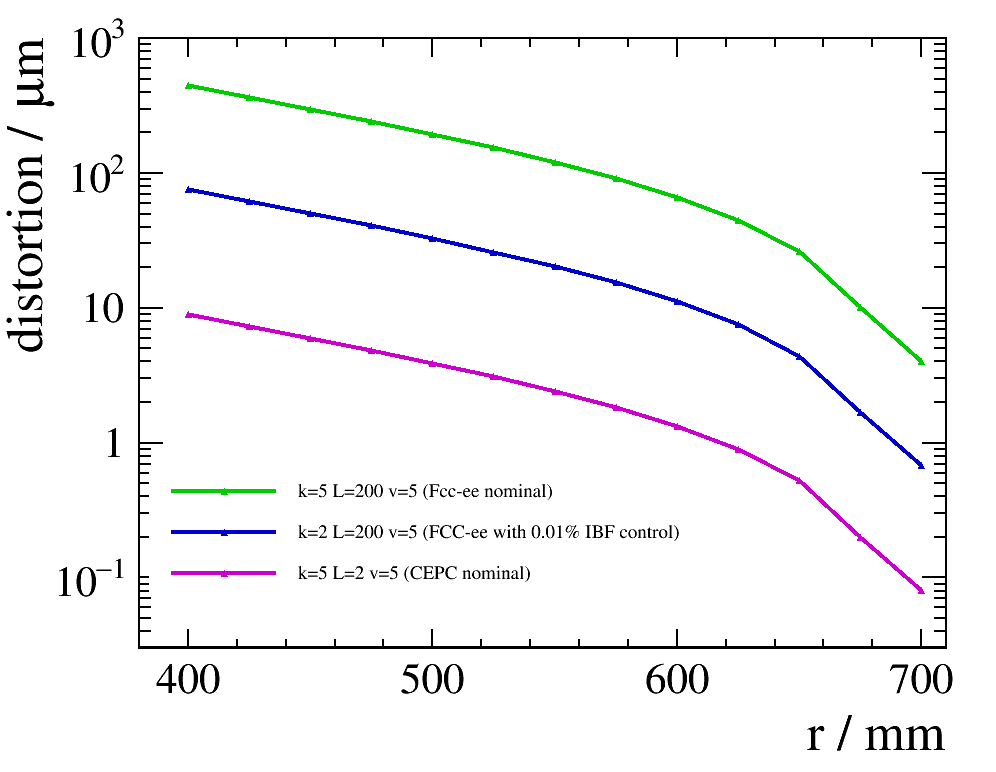}
        \vspace*{-0.5cm}
    \end{center}
    \caption{ \small
        Distortion as a function of electron initial r position with different parameters. 
    }
    \label{fig:distortion}
\end{figure}

The maximal distortion as a function of $r$ can be calculated under different parameters is shown in Figure~\ref{fig:distortion}.
The maximal distortion with $L = 2$, $\kb= 5$ and $\vion=5$ is an order of $10\mum$. It suggests that the distortion is safety for CEPC benchmark parameters. 
However, in the worst scenario with $\kb=100$ and $\vion=5$ and $L = 100$, corresponding to the designed luminosity of \fccee, the maximal distortion can reach an order of $10^4\mum$. 

A few approaches can be taken to mitigate the distortion. If the ion back flow can be controlled from $\kb = 100$ to $\kb = 5$. Decreasing the TPC length and increasing the magnetic field also help to reduce the distortion. 

Moreover, the distortion is always along the direction of $\textbf{E}\times\textbf{B}$. Given a hit with definite position, the distortion can be corrected back.

The momentum resolution of tracks are mainly determined by the silicon detector placed at the inner most layer of \cepc\ conceptual detector, while TPC is mainly used for track finding. Combining the factors together, TPC is also feasible for \fccee.

\section{Conclusion}

Using an sample of 9 thousand fully simulated $\ZToqq$ events at center of mass energy of $91.2 \gev$, we studied the voxel occupancy and the local charge density of the \cepc\ TPC at \Z pole operation for future circular electron positron colliders, with an instant luminosity of $2\E{34}$ to $2\E{36} \cmsec$.

Given the fact that the beam bunch is evenly distributed along the accelerator circumference, the voxel occupancy is extremely low ($2\E{-5}$/$2\E{-7}$ for the inner most layer and $3\E{-6}$/$3\E{-8}$ for average) and poses no pressure for the TPC usage.

The distortion on TPC hit positions induced by the ion charges is estimated with dedicated program and calculation. At instant luminosity of $1\E{36}$ and an ionback flow control of percent level, the distortion can be as large as 10 mm at the inner most TPC layer at the \cepc\ conceptual detector geometry, which is two orders of magnitude
larger than the intrinsic TPC spatial resolution. 


A few approaches are proposed to reduce the effects caused by distortion:

\begin{itemize}
    \item Ion back flow control technology; the ion back flow should be controlled to per mille level, in other word, only 1-10 back flow ions is allowed for each primary
        ionization.
    \item Dedicated distortion correction algorithm, for the inner most layers, which should result in a mitigation of the hit position distortion by 1 order of magnitude.
    \item Adequate track finding algorithm that could link the TPC track fragments to vertex tracks at high efficiency and purity.
\end{itemize}	

Taking all of these approaches account, the distortion can be mitigated by approximately 2 orders of magnitude.
To conclude, the pad occupancy and distortion stress no pressure to \cepc\ and if the above items can be achieved, the usage of TPC is also a feasible option at \fccee.

The author thanks Keisuke Fujii, Serguei Ganjour for the code of calculating the distortion and
the discussions. This study was supported by National Key Programme for S\&T Research and Development (Grant NO.: 2016YFA0400400), the National Natural Science Foundation of China (Grant
No.: 11675197) and the CAS Hundred Talent Program (Y3515540U1).

\bibliography{mybib}
\end{document}